\documentclass[twocolumn,showpacs,preprintnumbers,amsmath,amssymb]{revtex4}

\usepackage{graphicx}
\usepackage{dcolumn}
\usepackage{bm}

\begin{document}

\preprint{APS/123-QED}

\title{Clusters in weighted macroeconomic networks : the EU case. \\Introducing the overlapping index of  GDP/capita fluctuation correlations}

\author{M. Gligor}
\email{mrgligor@yahoo.com}

\affiliation{National College ÒRoman VodaÓ Roman-5550, Neamt, Romania}

\author{M. Ausloos}
\email{Marcel.Ausloos@ulg.ac.be}

\affiliation{
GRAPES, Universit\'e de Li\`ege, B5 Sart-Tilman, B-4000 Li\`ege, Belgium
}

\date{09/07/2005}

\begin{abstract}
GDP/capita correlations are investigated in various time windows (TW), for the time
interval 1990-2005. The target group of countries is the set of 25 EU members, 15 till
2004 plus the 10 countries which joined EU later on.  The TW-means of the statistical correlation coefficients
are taken as the weights (links) of a fully connected network having the countries as
nodes. Thereafter we define and  introduce the {\it overlapping index} of weighted network nodes. A cluster structure of EU countries is derived from the statistically
relevant eigenvalues and eigenvectors of the adjacency matrix. 
This may be considered to yield some information about the structure, stability
and evolution of the EU country clusters in a macroeconomic sense.
\end{abstract}


\maketitle

\section {Introduction} 

It is of major interest in economy to extract as much as possible information from the sparse and noisy macroeconomic (ME) time series. Most ME indicators have a yearly or at most quarterly frequency. When a ME indicator time series has been produced for a very long time, strong evidence against stationarity alas arises \cite{Joe1}. Therefore the correlation patterns are usually investigated by moving a constant size time window (TW) with a constant step so that the whole time interval is scanned, somewhat averaging the correlations.
   
   The goal of the present paper is to investigate the weighted fully connected network of the N = 25 countries forming the European Union in 2005 (EU-25). The ties between countries are supposed to be proportional to the degree of similitude of the macroeconomic fluctuations referring to the GDP/capita annual rates of growth between 1990 and 2005. The countries are abbreviated according to The Roots Web Surname List (RSL)  \cite{codes} which uses 3 letters standardized abbreviations to designate countries and other regional locations. The World Bank database  \cite{wb}  is here used as data source  instead of the Penn World Tables \cite{pwt} on which some data is missing for several East-European countries. In this way, the investigated time span goes from 1990 to 2005.
   
   The system is represented by an evolving network, nodes being the countries; links are $weights$ (or GDP/capita fluctuations). In order to extract structures from the network, we average the time correlations in different windows, i.e. we assume that one can consider ''average degrees'', etc. The matrix-based method reveals the emergence of a number of  ''common factors'', through  the main eigenvectors (Kaiser criterion and Cattel scree test).
 The usual expectations defined by politicians or economists   through geographical connexions \cite{aaberge, angelini, mora} are observed. 

Moreover we introduce the country {\it overlapping hierarchy index }. Further study would be of interest in order to see its role within  the network global evolution as a function of hierarchy levels through correlations with other indices of studies.
 
\section{Data and methdology}

 The here below investigated ME indicators are the GDP/capita annual growth rates. Indeed, the GDP/capita is expected to reflect to the largest extent what A. Smith called, over two centuries ago, "the wealth of nations". In fact, it is expected to account both for the economic development and for the people well being. The target group of countries is composed of $M = 25$ countries: the 15 members of the European Union in 2004 (EU-15) and the 10 countries which joined the European community in 2005 (EU-10).  
  
   \subsection{Correlation coefficient}
   
The correlation coefficients $C_{ij}$ between two ME time series {$x_{i}$} and {$y_{j}$},\; $ i,j=1,...,N$, is calculated in the present work according to the (Pearson's) classical formula:

\begin{equation}
C_{ij}(t, T)= {\frac{<x_{i}  y_{j}>-<x_{i}>  <y_{j}>}{\sqrt{<x_{i}^{2} - <x_{i}>^{2}>  <y_{j}^{2}-<y_{j}>^{2}>}}}
\end{equation}

Each $C_{ij}$ is clearly a function both of the time window size $T$ and of the initial time (i.e. the "position" of the constant size time window on the scanned time interval). One has to note (or recall) that the correlation coefficients are not additive, i.e. an average of correlation coefficients in a number of samples does not represent an "average correlation" in all those samples. In cases when one needs to average correlations, the $C_{ij}$ 's first have to be converted into additive measures. For example, one may square the $C_{ij}$ 's before averaging, to obtain the so called  \textit{coefficients of determination} ($C_{ij}^{2}$) which are additive, or one can convert the $C_{ij}$ 's into so-called \textit{Fisher z values}, which are also additive \cite{hays}. The former approach is used here below, so that the average correlations are calculated as:

\begin{equation}
\hat{C}_{i,j}(T)=\left[\frac{1}{\nu} \sum^{k+T}_{t=k} C^{2}_{ij}(t)\right]^{1/2} , 
\\ k=0, 1, \ldots,N-T, 
\end{equation}  

\noindent where $N$ is the total number of points (the time span), $T$ is the time window size used for the analysis, $\nu=N-T+1$ and $t$ is a discrete counter variable.

The weight of the connection between $i$ and $j$ reflects the strength of correlations between the two agents;  it can be simply expressed as:

\begin{equation}
w_{ij}(T)= \hat{C}_{i,j}(T)
\end{equation}

\noindent fulfilling the obvious relations: $0 \leq w_{ij} \leq 1$ ; $w_{ij} = w_{ji}$ and $w_{ij} = 1$ for $i = j$.

Thus an adjacency matrix with elements $w_{ij}(T)$ can be defined.
The cumulative distribution function (CDF)  of the weights is given in Fig. 1 for different  (4) time windows. One can see a remarkable sharpening   of the CDF as a function of decreasing  TW.

\begin{figure}
\centering
\includegraphics[height=10cm,width=10cm]{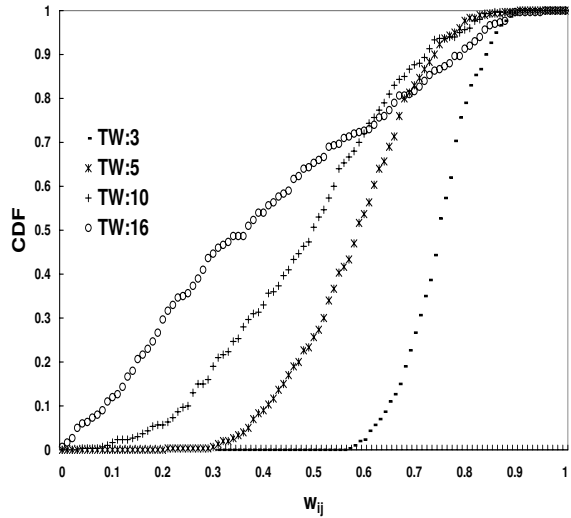}
\caption{\label{MG-Fig1} The cumulative distribution function (CDF)  of the weights set {$w_{ij}$} for four different time window sizes.}
\end{figure}

\begin{figure}
\centering
\includegraphics[height=10cm,width=10cm]{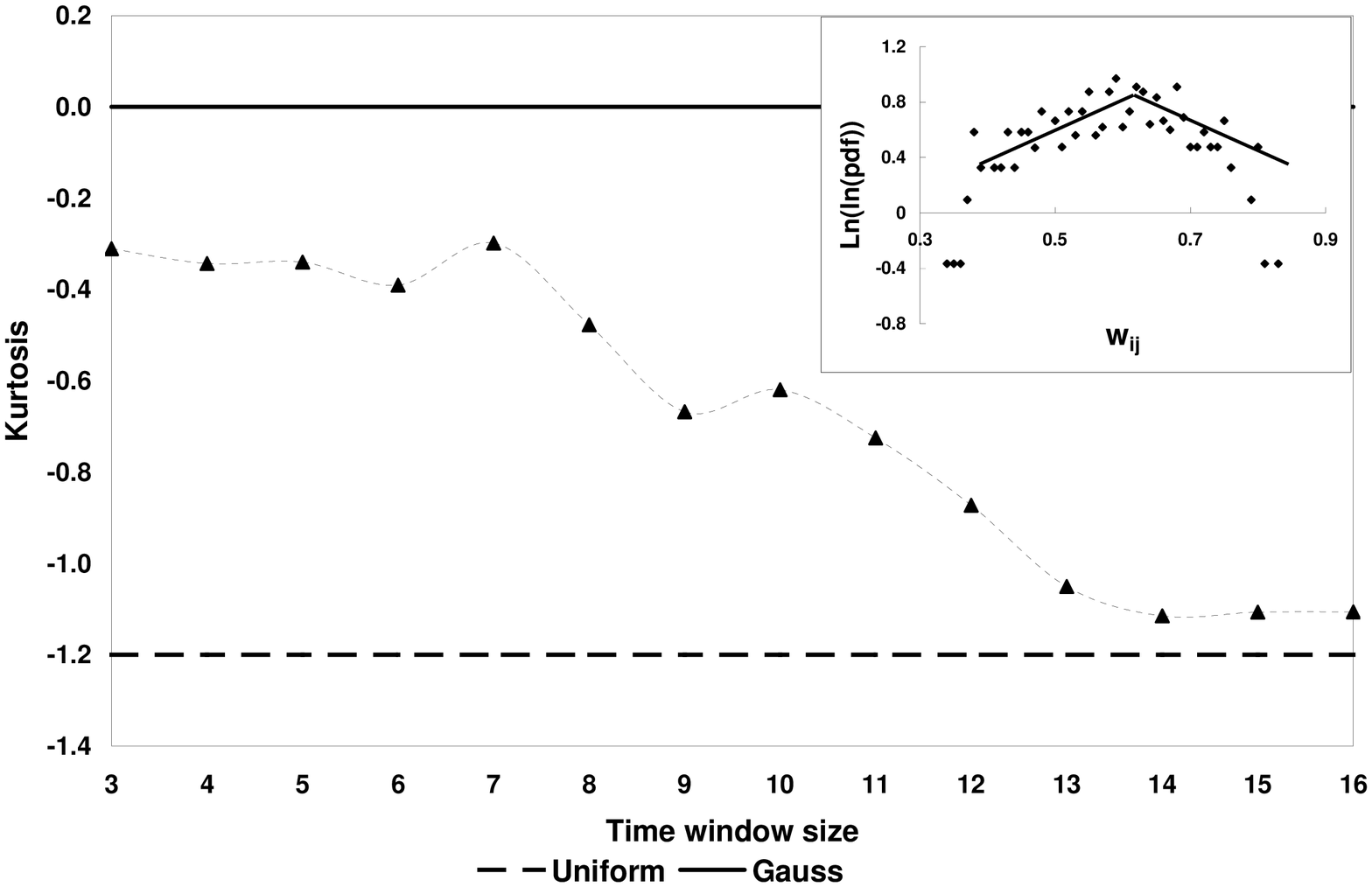}
\caption{\label{MG-Fig2} The kurtosis of the weights set {$w_{ij}$} versus the time window size. Inset: the double logarithm of the $w_{ij}$'s probability density function for 5 years time window size. The thick line has a  $\pm 2$ slope, corresponding to the Gaussian distribution.}
\end{figure}

\begin{figure}
\centering
\includegraphics[height=10cm,width=10cm]{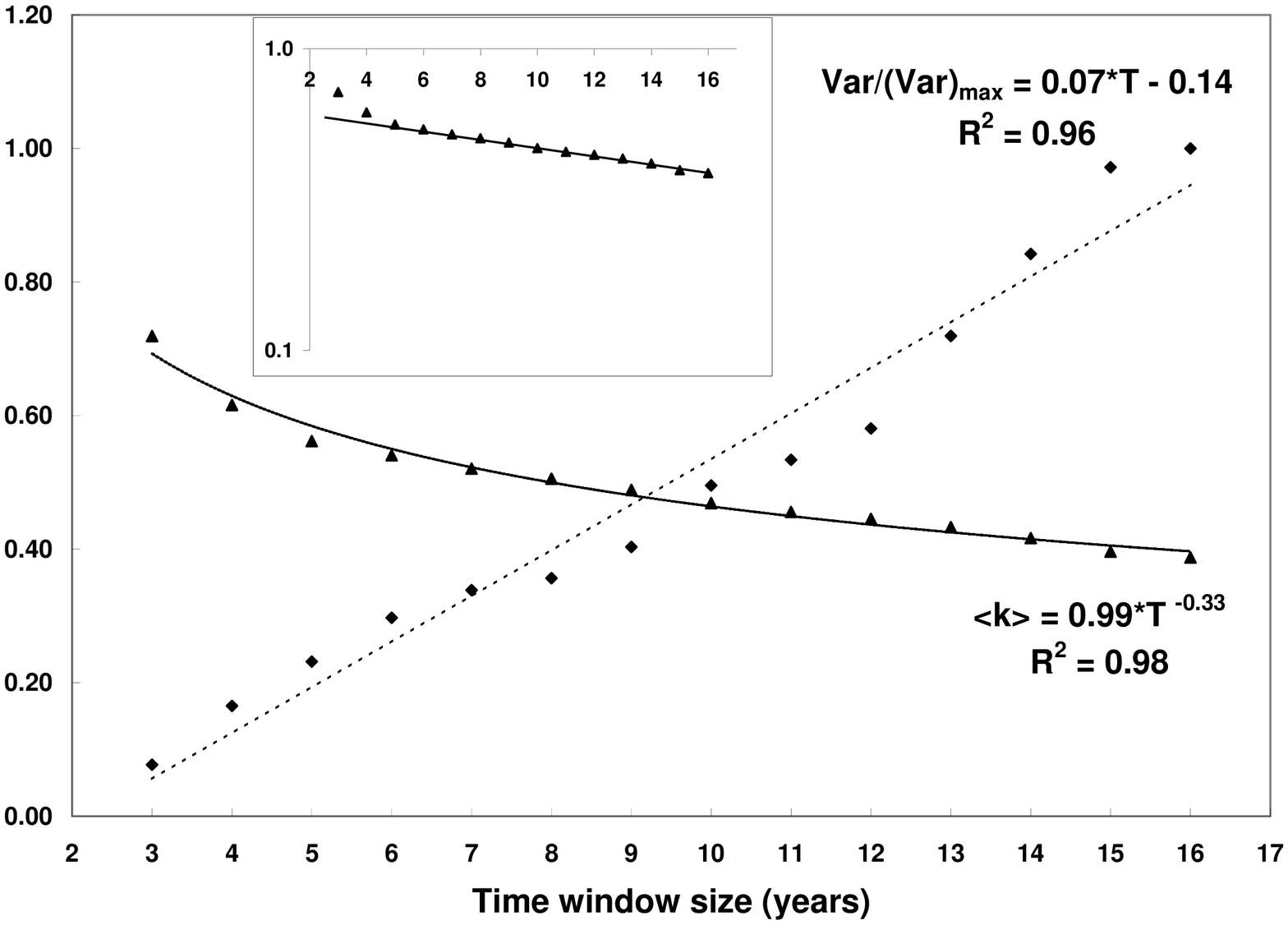}
\caption{\label{MG-Fig3} The average degree $<k>$ and the variance of the weights set {$w_{ij}$} in the EU-25 network versus the time window size. Variance is normalised to its maximal value. Inset: $<k>$ versus $T$ in log-plot for emphasizing the inverse cubic root law.}
\end{figure}

These changes of the distribution shape can also be pointed out through the kurtosis ($K$) variation with the time window size (Fig. 2). For the Gaussian distribution $K_{G} = 0$, while for the discrete uniform distribution of $m$ data ($m = 300$ here) it can be calculated \cite{hays} as:

\begin{equation}
K_{U}=-\frac{6}{5} \frac{m^{2}+1}{m^{2}-1} \approx -\frac{6}{5}
\end{equation}

It is found on Fig. 2 that the $K$ value shifts between the limit $K_{G}$ and $K_{U}$ indeed.

\subsection{Weighted network characteristics}  

\par Networks are characterized by various parameters. For instance, the vertex degree is the total number of vertex connections. It may be generalised in a weighted network \cite{albert, barrat, newman1, newman3, barabasi, pastsat} as:

\begin{equation}
k_{i}=\sum^{M}_{\substack {j=1 \\ j\neq i}}w_{ij}
\end{equation}

\noindent Thus, the average degree in the network is:

\begin{equation}
<k> = \frac{1}{M} \sum^{M}_{i=1}\sum^{M}_{\substack{j=1 \\ j\neq i}}w_{ij}
\end{equation}

as shown on Fig. 3. Notice a cubic root law behavior. 

Another characteristic quantity describes the number of triangles in the network indicating some correlations. In the literature, there have been several ways to evaluate assortative correlations, such as the assortativity coefficient introduced by Newman \cite{newman1} that is the Pearson correlation coefficient of the degrees at either ends of an edge. Nonetheless, all of them focus on local degree-correlations between two connected nodes. 

Here below we   introduce an \textit{overlapping index } $O_{ij}$ in order to indicate/find some hierarchy of clusters on a network.

First consider a non-weighted network consisting of $M$ vertices; let the number $N_{ij}$ measure the common number of neighbours of $i$ and $j$. We impose that  $O_{ij}$ must satisfy the following properties:

(1) $O_{ij} = 0\ \Leftrightarrow\ N_{ij} = 0$ (fully disconnected, or "tree-like" network);

(2) $O_{ij} = 1\ \forall i \neq j$ in a fully connected network, where $N_{ij} = M-2$ ; $k_i = k_j = M-1$;

(3) $0 < O_{ij} < 1$,\  otherwise;

(4) $O_{ij} \propto\  N_{ij}$ \ and \ $O_{ij} \propto\  <k>_{ij}\equiv  (k_{i} + k_{j})/2$

\noindent A quantity satisfying all these conditions (1)-(4) can be defined as:

\begin{equation}
O_{ij}=\frac{N_{ij} (k_{i} + k_{j})}{2(M-1)(M-2)}\ , \  i\neq j
\end{equation}

For a weighted network, Eq.(7) may be generalised as:

\begin{equation}
O_{ij}= A\sum^{M}_{\substack {l=1 \\ l\neq i,j}}(w_{il}+w_{jl})
\\
\left(\sum^{M}_{\substack {p=1 \\ p\neq i}} w_{ip} + \sum^{M}_{\substack {q=1 \\ q\neq j}} w_{jq} \right)  
\end{equation}

\noindent where $A = 1/[2(M-1)(M-2)]$ and $i\neq j$.  

One can easily see that $0 < O_{ij} < 1$, and $O_{ij} =1$ only for all $w_{ij} = 1$, i.e. fully connected non-weighted network. However, for a weighted network, $O_{ij}$ can never be zero.

Each overlapping index  is thus computed for each EU-25 country using the adjacency matrix defined in Eq. 3.  A country average overlapping index $<O_{i}>$ can be next assigned to each country, i.e. dividing the sum of its overlapping indices by its number of neighbours:

\begin{equation}
<O_{i}> = \frac{1}{M-1} \sum^{M}_{j=1} O_{ij}
\end{equation}

The results are shown in Table I. Due to lack of space, the reader is invited to draw some interesting conclusion by mere observation.

\begin{table}

\caption{Average overlapping index of each EU-25 country in decreasing order; 2 decimals have been used only}
\smallskip
\begin{footnotesize}
\begin{center}
\begin{tabular}{c c c c c c c c c c}

\\\hline 
\\SWE & 0.38 & SVK &	0.37 &	AUT &	0.35 &	POL &	0.34 &	LTU &	0.32\\
\\GER & 0.37 &	BEL &	0.36 &	FIN &	0.35 &	MLT &	0.33 &	LVA &	0.31\\
\\FRA &	0.37 &	IRL &	0.36 &	PRT &	0.35 &	GRC &	0.33 &	CZE &	0.31\\
\\DNK &	0.37 &	LUX &	0.36 &	NLD &	0.35 &	CYP &	0.32 &	EST &	0.30\\
\\HUN &	0.37 &	ESP &	0.35 &	ITA &	0.35 &	SVN &	0.32 &	GBR &	0.29\\
\\\hline

\end{tabular}
\end{center}
\end{footnotesize}
\end{table}

 \subsection{Clusters}

\begin{figure}
\centering
\includegraphics[height=10cm,width=10cm]{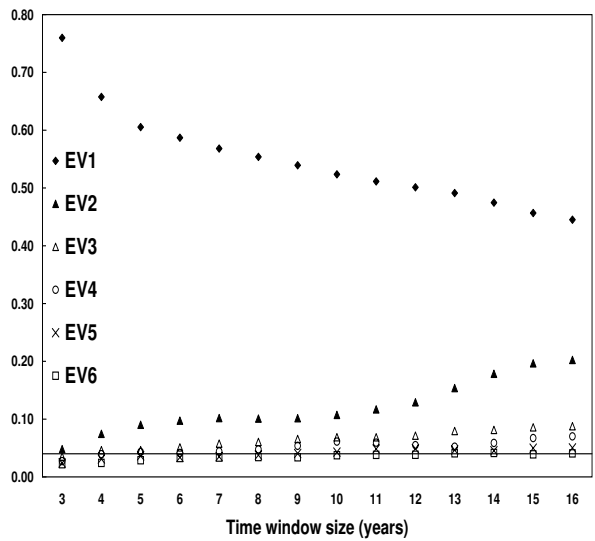}
\caption{\label{MG-Fig4} The six largest eigenvalues (EVZ;  Z = 1,6) of the adjacency matrix [$w_{ij}$] for the EU-25 weighted network versus the moving time window size. The eigenvalues are normalized to the correlation matrix size ($M = 25$), thus the vertical axis may be read as a fractional contribution to the total variance.}
\end{figure}

Finally let us question  whether clusters exist. This is done through the study of the eigenvalues and eigenvectors of the correlation  matrix defined here above.
Let us recall firstly that the eigenvalues can be interpreted as the proportion of variance explained by each canonical correlation relating two sets of variables. There will be as many eigenvalues as there are canonical correlations (roots), and each successive eigenvalue will be smaller than the last since each successive root will explain less and less of the data. In factor analysis, the eigenvectors of a correlation matrix correspond to factors, and eigenvalues to factor loadings. The observable random variables are modeled as linear combinations of the factors, plus the ÒerrorÓ terms. 

\begin{figure}
\centering
\includegraphics[height=10cm,width=10cm]{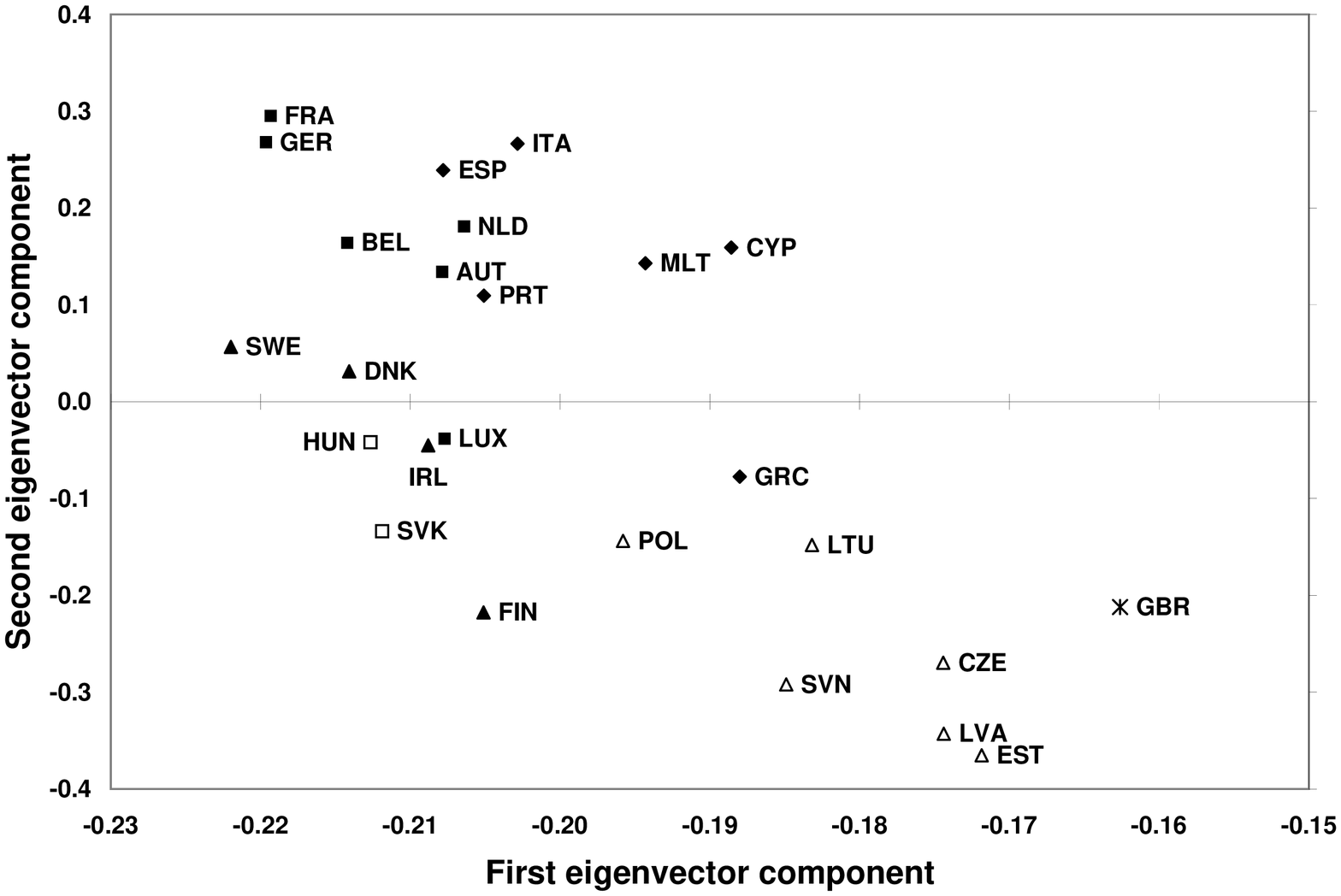}
\caption{\label{MG-Fig5} The cluster-like structure of the EU-25 countries according to the GDP/capita rates of growth. The country coordinates are the corresponding eigenvector components of the EU-25 weighted network adjacency matrix [$w_{ij}$].}
\end{figure}

   Having a measure of how much variance each successive factor extracts, one can call the question of how many factors to retain. By its nature this is somehow an arbitrary decision. However, there are some guidelines that are commonly used, and that, in practice, seem to yield the best results. Firstly, we can retain only factors with eigenvalues greater than 1. In essence this is like saying that, unless a factor extracts at least as much as the equivalent of one original variable, one has to drop it. This criterion, firstly proposed by Kaiser \cite{kaiser}, is probably the one most widely used. A graphical method is the ÒscreeÓ test first proposed by Cattell \cite{cattell}. Plotting the eigenvalues in a simple line plot, one has to find the place where the smooth decrease of eigenvalues appears to level off to the right of the plot. To the right of this point, presumably, one finds only Òfactorial screeÓ (ÒscreeÓ is the geological term referring to the debris which collects on the lower part of a rocky slope). Both criteria have been studied in detail \cite{kaiser, cattell, browne}. By generating random data based on a particular number of factors it was found that the first method (Kaiser criterion) sometimes retains too many factors, while the second technique (scree test) sometimes retains too few; However, both methods were found remarkably convergent when the number of common factors is not too large \cite{browne}. The above considerations explain why the first eigenvectors (i.e. the ones corresponding to the largest eigenvalues) are generally considered as carrying the useful information. 
   
   \begin{figure}
\centering
\includegraphics[height=10cm,width=10cm]{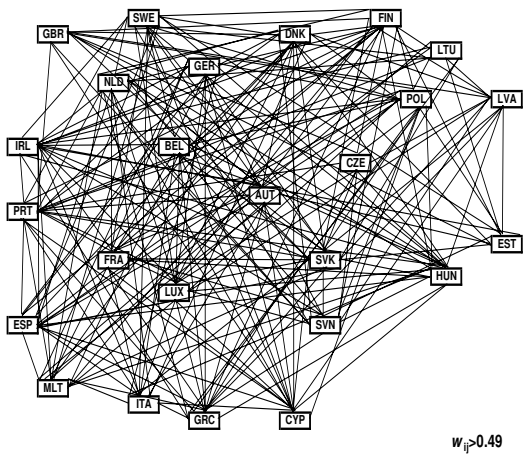}
\caption{\label{MG-Fig6} The EU-25 weighted network for the weights greater than the threshold  value $w_{T} = 0.49$}
\end{figure}

\begin{figure}
\centering
\includegraphics[height=10cm,width=10cm]{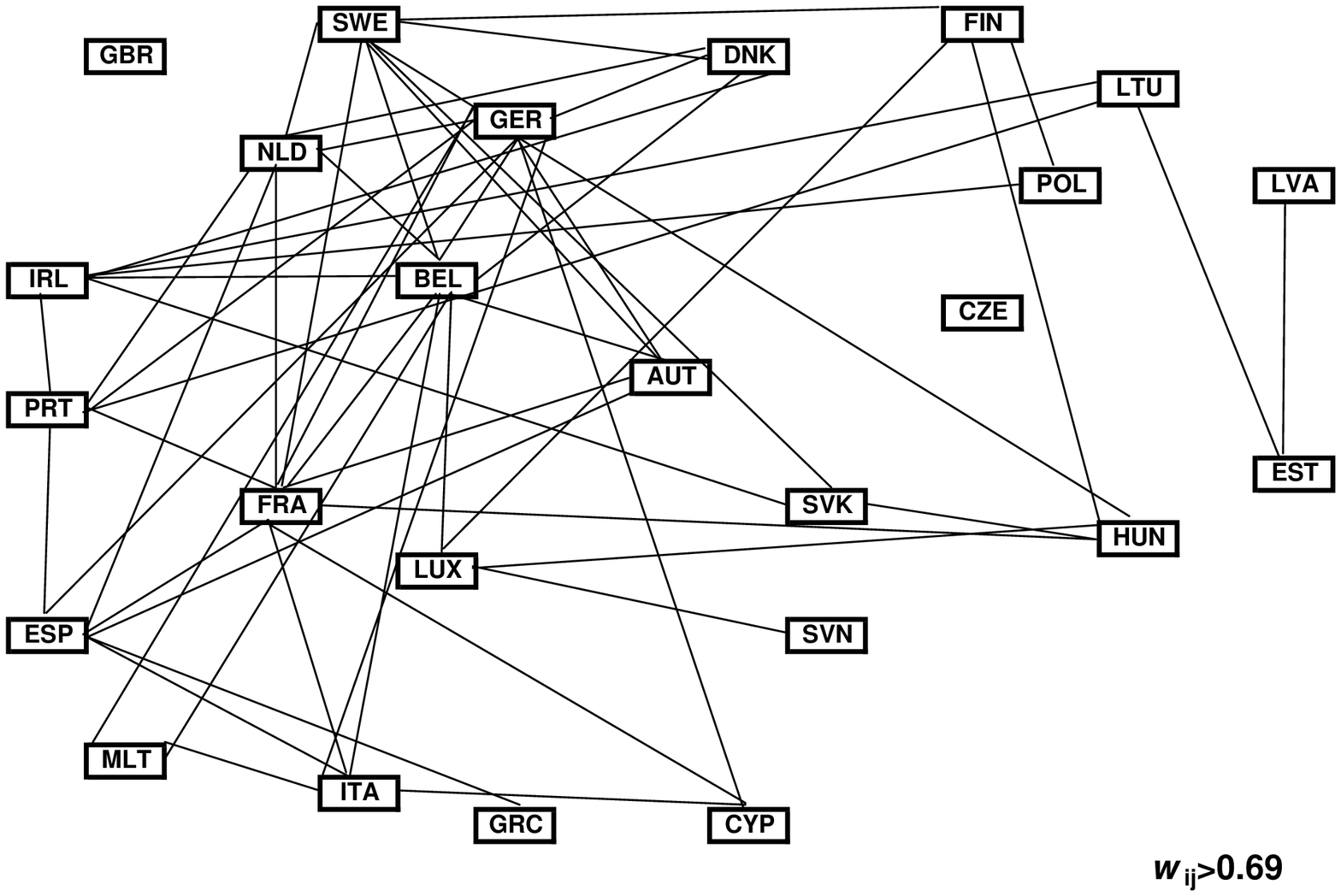}
\caption{\label{MG-Fig7} The EU-25 weighted network for the weights greater than the threshold  value $w_{T} = 0.69$}
\end{figure}

\begin{figure}
\centering
\includegraphics[height=10cm,width=10cm]{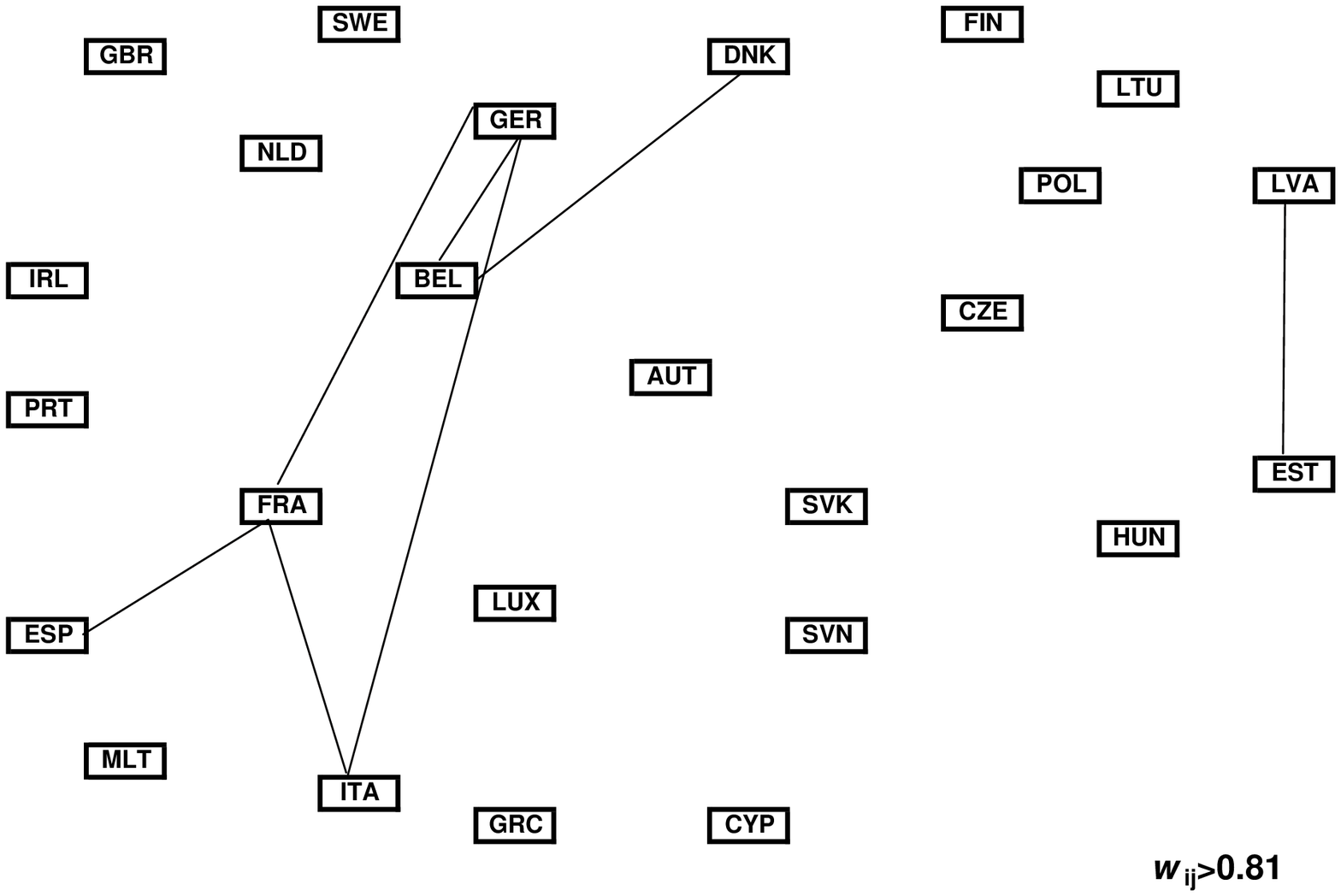}
\caption{\label{MG-Fig8} The EU-25 weighted network for the weights greater than the threshold  value $w_{T} = 0.81$}
\end{figure}

   The six largest eigenvalues (EVZ;  Z = 1,6) of the adjacency matrix [$w_{ij}$] for the EU-25 weighted network  are shown as a function of the moving time window size on Fig. 4. For the display the eigenvalues are normalized to the correlation matrix size ($M = 25$), thus the vertical axis may be read as a fractional contribution to the total variance. ''Obviously'', only the first two eigenvalues seem to be of major interest. Thus
a clustering scheme of the EU-25 countries can be constructed on the structure of the corresponding eigenvectors; the EU-25 countries are positioned  with respect to these vectors according to their coordinates in Fig. 5.

\section{Conclusion}

Complex networks have become an active field  of research in physics \cite{albert, barrat, newman1, newman3, barabasi, pastsat}.
These systems are usually composed of a large number of internal components (the nodes and links), and describe a wide variety 
of systems of high  intellectual  and technological importance. Relevant questions pertain to the critical dynamics of properties, not only on the network, but also for the network structure itself. The occurrence of structural  (''order-disorder'') transitions has been rarely studied.  Here we have reported on the structure of a 25 node cluster made of the EU countries linked together through weighted links representing the correlations between the GDP/capita. We have average those links in various time windows. For such weighted networks one can define a ''hierarchy overlapping coefficient''. Clusters in the network appear when filtering the weights through this coefficient.
 
Thus the adjacency matrix (eigenvalues, eigenvectors) with the Kaiser criterion allows to well observe
networks/clusters of countries. Notice the strong relationship between the average overlapping index (Table 1) hierarchy and the position of the countries derived from the correlation matrix main eigenvalues/vectors. 

Finally for any politically-economically minded observer we show (Figs. 6-8) the EU-25 network at various weight levels, i.e. eliminating links below various threshold  values.

In summary, some quantitative way to describe the structure, hierarchy
and evolution of the EU countries in a macroeconomic sense has been presented.

\begin{acknowledgments} MG would like to thank Francqui Foundation for financial support. MA acknowledges support from
Actions de recherche concert\'ees (ARC) (ARC 02/07-293)  from Communaut\'e
fran\c{c}aise de Belgique, Direction de la Recherche scientifique. Both authors acknowledge support from the European program  COST P10 {\it Physics of Risk} for its intense motivation.   \end{acknowledgments}

\end{document}